\documentclass[lettersize,journal]{IEEEtran}
\usepackage{amsmath,amsfonts}
\usepackage{algorithmic}
\usepackage{array}

\newif\ifarxiv 
\arxivtrue

\usepackage[font=small]{caption}
\usepackage{subcaption}

\usepackage{textcomp}
\usepackage{stfloats}
\usepackage{url}
\usepackage{verbatim}
\usepackage{graphicx}
\def\BibTeX{{\rm B\kern-.05em{\sc i\kern-.025em b}\kern-.08em
    T\kern-.1667em\lower.7ex\hbox{E}\kern-.125emX}}
\usepackage{balance}
\usepackage[dvipsnames,table,xcdraw]{xcolor}

\usepackage[acronym]{glossaries}
\usepackage{import}

\makeglossaries

\newacronym{2d}{2D}{two-dimensional}
\newacronym{3d}{3D}{three-dimensional}
\newacronym{3gpp}{3GPP}{3rd generation partnership project}
\newacronym{4g}{4G}{fourth-generation}
\newacronym{5g}{5G}{fifth-generation}
\newacronym{6dof}{6DoF}{six degrees of freedom}
\newacronym{6g}{6G}{sixth-generation}
\newacronym{abp}{ABP}{authentication by personalisation}
\newacronym{ac}{AC}{alternating current}
\newacronym{aclr}{ACLR}{adjacent channel leakage ratio}
\newacronym{adc}{ADC}{analog-to-digital converter}
\newacronym{adr}{ADR}{adaptive data rate}
\newacronym{afe}{AFE}{analog front end}
\newacronym{ag}{AG}{array gain}
\newacronym{ai}{AI}{artificial intelligence}
\newacronym{am}{AM}{amplitude modulation}
\newacronym{amam}{AM/AM}{amplitude modulation to amplitude modulation}
\newacronym{amc}{AMC}{automatic modulation classification}
\newacronym{amp}{AMP}{approximate message passing}
\newacronym{ampm}{AM/PM}{amplitude modulation to phase modulation}
\newacronym{aoa}{AOA}{angle-of-arrival}
\newacronym{aod}{AOD}{angle-of-departure}
\newacronym{ap}{AP}{access point}
\newacronym{api}{API}{application program interface}
\newacronym{apu}{APU}{access point unit}
\newacronym{ar}{AR}{augmented reality}
\newacronym{arima}{ARIMA}{Auto-Regressive Integrated Moving Average}
\newacronym{arp}{ARP}{Antenna Reference Point}
\newacronym{asic}{ASIC}{application specific integrated circuit}
\newacronym{ask}{ASK}{amplitude-shift keying}
\newacronym{auv}{AUV}{autonomous underwater vehicle}
\newacronym{awgn}{AWGN}{additive white Gaussian noise}
\newacronym{baw}{BAW}{bulk acoustic wave}
\newacronym{bb}{BB}{base-band}
\newacronym{bc}{BC}{backscatter communication}
\newacronym{bd}{BD}{backscatter device}
\newacronym{be}{BE}{Belgium}
\newacronym{ber}{BER}{bit error rate}
\newacronym{bf}{BF}{beamforming}
\newacronym{bldc}{BLDC}{brushless DC}
\newacronym{bom}{BOM}{bill of materials}
\newacronym{bpsk}{BPSK}{binary phase-shift keying}
\newacronym{bs}{BS}{base station}
\newacronym{bu}{BU}{booster unit}
\newacronym{bw}{BW}{bandwidth}
\newacronym{ca}{CA}{carrier emitter}
\newacronym{cad}{CAD}{channel activity detection}
\newacronym{cars}{CARS}{calibration reference signal}
\newacronym{cbm}{CBM}{condition based maintenance}
\newacronym{cc}{CC}{constant current}
\newacronym{ccdf}{CCDF}{complementary cumulative distribution function}
\newacronym{ccnn}{CCNN}{circular convolutional neural network}
\newacronym{ccs}{CCS}{correlative channel sounder}
\newacronym{cdf}{CDF}{cumulative distribution function}
\newacronym{cdrx}{CDRX}{connected mode DRX}
\newacronym{ce}{CE}{coverage enhancement}
\newacronym{ced}{CED}{cumulative energy density}
\newacronym{cf}{CF}{cell-free}
\newacronym{cfmmimo}{CF-mMIMO}{cell-free massive MIMO}
\newacronym{cfo}{CFO}{carrier frequency offset}
\newacronym{cir}{CIR}{channel impulse response}
\newacronym{cla}{CLA}{closed-loop approach}
\newacronym{cmos}{CMOS}{complementary metal oxide semiconductor}
\newacronym{cost}{COST}{commercial off-the-shelf}
\newacronym{cots}{COTS}{commercial off-the-shelf}
\newacronym{cp}{CP}{cyclic prefix}
\newacronym{cpt}{CPT}{capacitive power transfer}
\newacronym{cpu}{CPU}{central-processing unit}
\newacronym{cqi}{CQI}{channel quality indicator}
\newacronym{cr}{CR}{coding rate}
\newacronym{crc}{CRC}{cyclic redundancy check}
\newacronym{crlb}{CRLB}{Cram\'er-Rao lower bound}
\newacronym{crs}{CRS}{cell reference signal}
\newacronym{cs}{CS}{compressed sensing}
\newacronym{cse}{CSE}{channel state estimation}
\newacronym{csi}{CSI}{channel state information}
\newacronym{csp}{CSP}{contact service point}
\newacronym{css}{CSS}{chirp spread spectrum}
\newacronym{cu}{CU}{central unit}
\newacronym{cv}{CV}{constant voltage}
\newacronym{cw}{CW}{continuous wave}
\newacronym{d2d}{D2D}{device-to-device}
\newacronym{dab}{DAB}{Digital Audio Broadcasting}
\newacronym{dac}{DAC}{digital-to-analog converter}
\newacronym{daq}{DAQ}{data acquisition system}
\newacronym{das}{DAS}{distributed antenna systems}
\newacronym{dc}{DC}{direct current}
\newacronym{dcc}{DCC}{dynamic cooperation clustering}
\newacronym{de}{DE}{drain efficiency}
\newacronym{dft}{DFT}{discrete Fourier transform}
\newacronym{dl}{DL}{downlink}
\newacronym{dlc}{DLC}{Distributed Laser Charging}
\newacronym{dli}{DLI}{direct link interference}
\newacronym{dm}{DM}{diffuse multipath}
\newacronym{dma}{DMA}{Direct Memory Access}
\newacronym{dmc}{DMC}{diffuse multipath component}
\newacronym{dmimo}{D-MIMO}{distributed MIMO}
\newacronym{doa}{DOA}{direction-of-arrival}
\newacronym{dpd}{DPD}{digital pre-distortion}
\newacronym{dpdk}{DPDK}{Data Plane Development Kit}
\newacronym{dram}{DRAM}{dynamic random-access memory}
\newacronym{drcs}{$\Delta$RCS}{differential-radar cross section}
\newacronym{drx}{DRX}{Discontinuous Reception Mode}
\newacronym{dsl}{DSL}{digital subscriber line}
\newacronym{dsp}{DSP}{digital signal processing}
\newacronym{dss}{DSS}{Dataset Storage Standard}
\newacronym{duc}{DUC}{digital up-converter}
\newacronym{dvb}{DVB}{Digital Video Broadcasting}
\newacronym{e2e}{E2E}{end-to-end}
\newacronym{easa}{EASA}{European Union Aviation Safety Agency}
\newacronym{ec}{EC}{European Commission}
\newacronym{ecdf}{eCDF}{empirical cumulative distribution function}
\newacronym{ecsp}{ECSP}{edge computing service point}
\newacronym{edlc}{EDLC}{electrostatic double-layer capacitors}
\newacronym{edrx}{eDRX}{Extended Discontinuous Reception Mode}
\newacronym{ee}{EE}{energy efficiency}
\newacronym{egprs}{EGPRS}{Enhanced Data Rates for GSM Evolution}
\newacronym{eh}{EH}{energy harvesting}
\newacronym{eirp}{EIRP}{equivalent isotropically radiated power}
\newacronym{em}{EM}{electromagnetic}
\newacronym{embb}{eMBB}{enhanced Mobile Broadband}
\newacronym{en}{EN}{energy neutral}
\newacronym{end}{END}{energy neutral device}
\newacronym{enob}{ENOB}{effective number of bits}
\newacronym{eol}{EoL}{end of life}
\newacronym{epu}{EPU}{edge processing unit}
\newacronym{er}{ER}{Energy Receiver}
\newacronym{erp}{ERP}{equivalent radiated power}
\newacronym[plural=ESCs,firstplural=electronic speed controllers (ESCs)]{esc}{ESC}{electronic speed control}
\newacronym{esd}{ESD}{electrostatic discharge}
\newacronym{esl}{ESL}{electronic shelf label}
\newacronym{esr}{ESR}{equivalent series resistance}
\newacronym{et}{ET}{Energy Transmitter}
\newacronym{etsi}{ETSI}{European Telecommunications Standards Institute}
\newacronym{evd}{EVD}{eigenvalue decomposition}
\newacronym{evm}{EVM}{Error Vector Magnitude}
\newacronym{ewlb}{eWLB}{Embedded Wafer Level Ball Grid Array}
\newacronym{fa}{FA}{federation anchor}
\newacronym{fair}{FAIR}{Findability, Accessibility, Interoperability, and Reuse of digital assets}
\newacronym{fd}{FD}{front-haul distance}
\newacronym{fde}{FDE}{frequency domain equalizer}
\newacronym{fdm}{FDM}{frequency-division multiplexing}
\newacronym{fdma}{FDMA}{frequency division multiple access}
\newacronym{fem}{FEM}{finite element analysis}
\newacronym{fembb}{feMBB}{further enhanced mobile broadband}
\newacronym{fft}{FFT}{fast Fourier transform}
\newacronym{fh}{FH}{fronthaul}
\newacronym{fifo}{FIFO}{First In, First Out}
\newacronym{fim}{FIM}{Fisher information matrix}
\newacronym{fits}{FITS}{Flexible Image Transport System}
\newacronym{fom}{FoM}{figure of merit}
\newacronym{fpga}{FPGA}{field-programmable gate array}
\newacronym{fsk}{FSK}{frequency shift keying}
\newacronym{fss}{FSS}{frequency selective surface}
\newacronym{gb}{GB}{grant-based}
\newacronym{gf}{GF}{grant-free}
\newacronym{gmsk}{GMSK}{Gaussian minimum-shift keying}
\newacronym{gnb}{gNB}{Next Generation Node B}
\newacronym{gni}{GNI}{gross national income}
\newacronym{gnn}{GNN}{graph neural network}
\newacronym{gnss}{GNSS}{global navigation satellite system}
\newacronym{gpclk}{GPCLK}{general purpose clock}
\newacronym{gpl}{GPL}{GNU General Public License}
\newacronym{gprs}{GPRS}{General Packet Radio Services}
\newacronym{gps}{GPS}{Global Positioning System}
\newacronym{gpu}{GPU}{graphical processing unit}
\newacronym{grc}{GRC}{GNU Radio Companion}
\newacronym{gscm}{GSCM}{geometry‐based stochastic model}
\newacronym{gsm}{GSM}{Global System for Mobile Communications}  %
\newacronym{gwp}{GWP}{Global Warming Potential}
\newacronym{harq}{HARQ}{hybrid automatic repeat request}
\newacronym{hat}{HAT}{hardware attached on top}
\newacronym{hcs}{HCS}{human-centric services}
\newacronym{hdf5}{HDF5}{Hierarchical Data Format version 5}
\newacronym{hfss}{HFSS}{High Frequency Simulator Software}
\newacronym{hmd}{HMD}{head-mounted display}
\newacronym{hpbm}{HPBM}{half power beam width}
\newacronym{HyMPRo}{HyMPRo}{Hybrid Multi-Path Routing algorithm}
\newacronym{i.i.d.}{i.i.d.}{independent and identically distributed}
\newacronym{ib}{IB}{in-band}
\newacronym{ibo}{IBO}{input back-off}
\newacronym{ic}{IC}{integrated circuit}
\newacronym{ici}{ICI}{intercarrier interference}
\newacronym{id}{ID}{information decoding}
\newacronym{idft}{IDFT}{inverse discrete Fourier transform}
\newacronym{if}{IF}{intermediate-frequency}
\newacronym{iid}{i.i.d.}{independently and identically distributed}
\newacronym{iis}{IIS}{integrated information system}
\newacronym{im}{IM}{intermodulation}
\newacronym{imu}{IMU}{inertial measurement unit}
\newacronym{io}{IO}{input/output}
\newacronym{iot}{IoT}{Internet of Things}
\newacronym{ipt}{IPT}{inductive power transfer}
\newacronym{ipy}{IPY}{Interventions per Year}
\newacronym{iq}{IQ}{in-phase and quadrature}
\newacronym{iqi}{IQI}{IQ imbalance}
\newacronym{ir}{IR}{infrared}
\newacronym{isi}{ISI}{intersymbol interference}
\newacronym{ism}{ISM}{industrial, scientific and medical}
\newacronym{isp}{ISP}{internet service provider}
\newacronym{jesd}{JESD}{Joint Electron Devices Engineering Council}
\newacronym{jfet}{JFET}{junction field effect transistor}
\newacronym{kpi}{KPI}{key performance indicator}
\newacronym{kvi}{KVI}{key value indicator}
\newacronym{larva}{LARVA}{LARge Virtual Array}
\newacronym{lca}{LCA}{life cycle assessment}
\newacronym{lco}{LCO}{lithium cobalt oxide}
\newacronym{ldo}{LDO}{Low-dropout voltage regulator}
\newacronym{ldpc}{LDPC}{low-density parity-check}
\newacronym{less}{LESS}{Low Energy Scheduler Solution}
\newacronym{lfp}{LFP}{lithium iron phosphate}
\newacronym{lic}{LIC}{lithium-ion capacitor}
\newacronym{lidar}{LiDAR}{light detection and ranging}
\newacronym{liion}{Li-ion}{lithium-ion}
\newacronym{lipo}{LiPo}{lithium polymer}
\newacronym{lis}{LIS}{large intelligent surface}
\newacronym{llh}{LLH}{log-likelihood}
\newacronym{lmmse}{LMMSE}{least minimum mean square error}
\newacronym{lmo}{LMO}{lithium ion manganese oxide}
\newacronym{lna}{LNA}{low-noise amplifier}
\newacronym{lo}{LO}{local oscillator}
\newacronym{lora}{LoRa}{long range}
\newacronym{lorawan}{LoRaWAN}{long-range wide-area network}
\newacronym{los}{LoS}{line-of-sight}
\newacronym{lp}{LP}{linear programming}
\newacronym{lpt}{LPT}{laser power transfer}
\newacronym{lpwa}{LPWA}{Low Power Wide Area}
\newacronym{lpwan}{LPWAN}{low-power wide-area network}
\newacronym{lqi}{LQI}{link quality indicator}
\newacronym{lrelu}{LReLU}{leaky rectified linear unit}
\newacronym{lrt}{LRT}{likelihood-ratio test}
\newacronym{ls}{LS}{least squares}
\newacronym{lsa}{LSA}{large synthetic array}
\newacronym{lsf}{LSF}{large-scale fading}
\newacronym{lsfc}{LSFC}{large-scale fading component}
\newacronym{lstm}{LSTM}{Long Short-Term Memory}
\newacronym{ltc}{LTC}{lithium thionyl chloride}
\newacronym{lte}{LTE}{Long Term Evolution}
\newacronym{lti}{LTI}{linear time-invariant}
\newacronym{lto}{LTO}{lithium titanate}
\newacronym{lusta}{LUSTA 5G }{Logistique mUltimodale Sécuritaire Téléopérée \& Autonome 5G}
\newacronym{m2m}{M2M}{machine to machine}
\newacronym{mac}{MAC}{Medium Access Control}
\newacronym{mate}{MATE}{millimeter-wave MIMO testbed}
\newacronym{mc}{MC}{Monte Carlo}
\newacronym{mcl}{MCL}{Maximum Coupling Loss}
\newacronym{mcs}{MCS}{modulation and coding scheme}
\newacronym{mcu}{MCU}{Microcontroller Unit}
\newacronym{mec}{MEC}{multi-access edge computing}
\newacronym{mems}{MEMS}{micro-electromechanical systems}
\newacronym{mf}{MF}{matched filter}
\newacronym{mimo}{MIMO}{multiple-input multiple-output}
\newacronym{miso}{MISO}{multiple-input single-output}
\newacronym{ml}{ML}{machine learning}
\newacronym{mlp}{MLP}{multilayer perceptron}
\newacronym{mmimo}{mMIMO}{massive MIMO}
\newacronym{mmse}{MMSE}{minimum mean square error}
\newacronym{mmtc}{mMTC}{massive machine-typed communication}
\newacronym{mmwave}{mmWave}{millimeter wave}
\newacronym{mn}{MN}{matching network}
\newacronym{mosfet}{MOSFET}{metal-oxide semiconductor field effect transistor}
\newacronym{mpc}{MPC}{multipath component}
\newacronym{mppt}{MPPT}{maximum power point tracking}
\newacronym{mr}{MR}{maximum ratio}
\newacronym{mrc}{MRC}{maximum ratio combining}
\newacronym{mrc_em}{MRC}{maximum ratio combining}
\newacronym{mrc_EM}{MRC}{Magnetic Resonance Coupling}
\newacronym{mrt}{MRT}{maximum ratio transmission}
\newacronym{mse}{MSE}{mean square error}
\newacronym{mtc}{MTC}{Machine-Type Communication}
\newacronym{multi-rat}{Multi-RAT}{multiple radio access technology}
\newacronym{music}{MUSIC}{MUltiple SIgnal Classification}
\newacronym{navauwall}{NAVAUWALL}{AUtomated NAVigation in WALLonia}
\newacronym{nb}{NB}{narrowband}
\newacronym{nbiot}{NB-IoT}{narrowband IoT}
\newacronym{nca}{NCA}{nickel cobalt aluminum}
\newacronym{netcdf}{NetCDF}{Network Common Data Form}
\newacronym{nfc}{NFC}{near-field communication}
\newacronym{nfv}{NFV}{network function virtualization}
\newacronym{ngmn}{NGMN}{Next Generation Mobile Networks }
\newacronym{ni}{NI}{National Instruments}
\newacronym{nicd}{NiCd}{nikkel cadmium}
\newacronym{nimh}{NiMH}{nikkel metal hydride}
\newacronym{nlos}{NLoS}{non-line-of-sight}
\newacronym{nmc}{NMC}{nickel manganese cobalt}
\newacronym{nmos}{nMOS}{n-channel metal-oxide semiconductor}
\newacronym{nn}{NN}{neural network}
\newacronym{nnls}{NNLS}{non-negative least squares}
\newacronym{noma}{NOMA}{non-orthogonal multiple access}
\newacronym{np}{NP}{Neyman-Pearson}
\newacronym{npbch}{NPBCH}{Narrowband Physical Broadcast Channel}
\newacronym{npss}{NPSS}{Narrow Band Primay Synchronization Signal}
\newacronym{nr}{NR}{New Radio}
\newacronym{nrs}{NRS}{Narrow Band Reference Signal}
\newacronym{nsss}{NSSS}{Narrowband Secondary Synchronization Signal}
\newacronym{ntp}{NTP}{network time protocol}
\newacronym{oai}{OAI}{OpenAirInterface} %
\newacronym{obw}{OBW}{occupied bandwidth}
\newacronym{ofdm}{OFDM}{orthogonal frequency-division multiplexing}
\newacronym{ofdma}{OFDMA}{orthogonal frequency-division multiple access}
\newacronym{ofdmim}{OFDM-IM}{OFDM with index modulation}
\newacronym{oob}{OOB}{out-of-band}
\newacronym{ook}{OOK}{on-off keying}
\newacronym{oran}{O-RAN}{open radio-access network}
\newacronym{os}{OS}{operating system}
\newacronym{ota}{OTA}{over-the-air}
\newacronym{otaa}{OTAA}{over-the-air authentication}
\newacronym{p1}{P1}{Phase 1}
\newacronym{p2}{P2}{Phase 2}
\newacronym{p2p}{P2P}{point-to-point}
\newacronym{pa}{PA}{power amplifier}
\newacronym{pae}{PAE}{power-added efficiency}
\newacronym{pana}{PanA}{Panel A}
\newacronym{panb}{PanB}{Panel B}
\newacronym{papr}{PAPR}{peak-to-average power ratio}
\newacronym{pc}{PC}{pilot count}
\newacronym{pcb}{PCB}{printed circuit board}
\newacronym{pcg}{PCG}{power consumption gain}
\newacronym{pcie}{PCIe}{Peripheral Component Interconnect Express}
\newacronym{pcsi}{PCSI}{perfect channel state information}
\newacronym{pd}{PD}{powered device}
\newacronym{pdcch}{PDCCH}{physical downlink control channel}
\newacronym{pdf}{PDF}{probability density function}
\newacronym{pdp}{PDP}{power delay profile}
\newacronym{pdsch}{PDSCH}{physical downlink shared channel}
\newacronym{peb}{PEB}{positioning error bound}
\newacronym{per}{PER}{packet error rate}
\newacronym{pg}{PG}{path gain}
\newacronym{pgd}{PGD}{proximal gradient descent}
\newacronym{phy}{PHY}{physical}
\newacronym{pl}{PL}{path loss}
\newacronym{pll}{PLL}{phase-locked loop}
\newacronym{pmf}{PMF}{polymer microwave fiber}
\newacronym{poe}{PoE}{power-over-Ethernet}
\newacronym{pps}{1PPS}{pulse per second}
\newacronym{prbs}{PRBs}{Physical Resource Blocks}
\newacronym{ps}{PS}{Processing System}
\newacronym{psd}{PSD}{power spectral density}
\newacronym{pse}{PSE}{power sourcing equipment}
\newacronym{psk}{PSK}{phase shift keying}
\newacronym{psm}{PSM}{power saving mode}
\newacronym{pss}{PSS}{primary synchronisation signal}
\newacronym{ptp}{PTP}{precision-time protocol}
\newacronym{ptrs}{PTRS}{Phase-Tracking Reference Signals}
\newacronym{ptw}{PTW}{paging time window}
\newacronym{pv}{PV}{photovoltaic}
\newacronym{pw}{PW}{planar wavefront}
\newacronym{pwm}{PWM}{pulse width modulation}
\newacronym{qam}{QAM}{quadrature amplitude modulation}
\newacronym{qos}{QoS}{quality-of-service}
\newacronym{qpsk}{QPSK}{quadrature phase-shift keying}
\newacronym{quadriga}{QuaDRiGa}{QUAsi Deterministic RadIo channel GenerAtor}
\newacronym{ra}{RA}{Random Access}
\newacronym{ram}{RAM}{random-access memory}
\newacronym{ran}{RAN}{radio access network}
\newacronym{rar}{RAR}{Random Access Response}
\newacronym{rat}{RAT}{radio access technology}
\newacronym{rb}{Rb}{Rubidium}
\newacronym{rbs}{RBS}{radio base station}
\newacronym{rbw}{RBW}{resolution bandwidth}
\newacronym{rcs}{RCS}{radar cross section}
\newacronym{re}{RE}{radio element}
\newacronym{relu}{ReLU}{rectified linear unit}
\newacronym{rf}{RF}{radio frequency}
\newacronym{rfeh}{RFEH}{radio frequency energy harvesting}
\newacronym{rfic}{RFIC}{radio-frequency integrated circuit}
\newacronym{rfid}{RFID}{radio frequency identification}
\newacronym{rfpt}{RFPT}{radio frequency power transfer}
\newacronym{rfsoc}{RFSoC}{Radio Frequency System-on-Chip}
\newacronym{rir}{RIR}{room impulse response}
\newacronym{ris}{RIS}{reflective intelligent surface}%
\newacronym{rllmtc}{RLLMTC}{reliable low latency machine type communication}
\newacronym{rls}{RLS}{recursive least squares}
\newacronym{rms}{RMS}{root-mean-square}
\newacronym{rmse}{RMSE}{root-mean-square error}
\newacronym{rmt}{RMT}{random matrix theory}
\newacronym{rof}{RoF}{radio-over-fiber}
\newacronym{ros}{ROS}{robot operating system}
\newacronym{rpi}{RPi}{Raspberry Pi}
\newacronym{rrc}{RRC}{Radio Resource Connection}
\newacronym{rreq}{RREQ}{route request packet}
\newacronym{rsrp}{RSRP}{Reference Signals Received Power}
\newacronym{rsrq}{RSRQ}{Reference Signal Received Quality}
\newacronym{rss}{RSS}{received signal strength}
\newacronym{rssi}{RSSI}{received signal strength indicator}
\newacronym{rtc}{RTC}{real time clock}
\newacronym{rtk}{RTK}{real time kinematics}
\newacronym{rts}{RTS}{ray tracing simulator}
\newacronym{ru}{RU}{Radio Unit}
\newacronym{rv}{RV}{random variable}
\newacronym{rw}{RW}{RadioWeaves}
\newacronym{rx}{RX}{receiver}
\newacronym{rzf}{RZF}{regularized zero forcing}
\newacronym{s-parameter}{S-parameter}{scattering parameter}
\newacronym{sa}{SA}{synchronization anchor}
\newacronym{sa5}{SA}{Stand Alone}
\newacronym{sar}{SAR}{specific absorption rate}
\newacronym{sbl}{SBL}{sparse Bayesian learning}
\newacronym{scfdma}{SCFDMA}{single-carrier frequency division multiple access}
\newacronym{sdg}{SDG}{Sustainable Development Goal}
\newacronym{sdm}{SDM}{sigma-delta modulator}
\newacronym{sdn}{SDN}{software-defined network}
\newacronym{sdof}{SDoF}{sigma-delta over fiber}
\newacronym{sdr}{SDR}{software-defined radio}
\newacronym{se}{SE}{spectral efficiency}
\newacronym{sei}{SEI}{specific emitter identification}
\newacronym{ser}{SER}{symbol-error rate}
\newacronym{sf}{SF}{spreading factor}
\newacronym{sfn}{SFN}{single frequency network}
\newacronym{sfp}{SFP}{small form-factor pluggable}
\newacronym{SigMF}{SigMF}{Signal Metadata Format}
\newacronym{simo}{SIMO}{single-input multiple-output}
\newacronym{sinr}{SINR}{signal-to-interference-plus-noise ratio}
\newacronym{siso}{SISO}{single-input single-output}
\newacronym{slam}{SLAM}{simultaneous localization and mapping}
\newacronym{slc}{SLC}{spatial leakage suppression}
\newacronym{slerp}{SLERP}{spherical linear interpolation}
\newacronym{sma}{SMA}{SubMiniature version A}
\newacronym{smc}{SMC}{specular multipath component}
\newacronym{smps}{SMPS}{switched mode power supply}
\newacronym{sndr}{SNDR}{signal-to-noise-and-distortion ratio}
\newacronym{snidr}{SNIDR}{signal-to-noise-and-interference-and-distortion ratio}
\newacronym{snr}{SNR}{signal-to-noise ratio}
\newacronym{soc}{SoC}{state of charge}
\newacronym{sram}{SRAM}{static random-access memory}
\newacronym{srd}{SRD}{short-range device}
\newacronym{srs}{SRS}{Sounding Reference Signal}
\newacronym{ssb}{SSB}{synchronisation signal block}
\newacronym{ssd}{SSD}{solid state drive}
\newacronym{ssq}{SSQ}{simulator sickness questionnaire}
\newacronym{steam}{STEAM}{science, technology, engineering, the arts, and mathematics}
\newacronym{svd}{SVD}{singular value decomposition}
\newacronym{sw}{SW}{spherical wavefront}
\newacronym{swipt}{SWIPT}{simultaneous wireless information and power transfer}
\newacronym{synce}{SyncE}{Synchronous Ethernet}
\newacronym{tau}{TAU}{tracking area update}
\newacronym{tcer}{TCER}{transported to consumed energy ratio}
\newacronym{tcp}{TCP}{Transmission Control Protocol}
\newacronym{tdd}{TDD}{time division duplexing}
\newacronym{tdoa}{TDOA}{time-difference-of-arrival}
\newacronym{thz}{THz}{Terahertz}
\newacronym{to}{TO}{timing offset}
\newacronym{toa}{TOA}{time-of-arrival}
\newacronym{tof}{ToF}{time-of-flight}
\newacronym{tosm}{TOSM}{through-open-short-match}
\newacronym{trl}{TRL}{technology readyness level}
\newacronym{trp}{TRP}{Transmission Reception Point}
\newacronym{tsn}{TSN}{time-sensitive networking}
\newacronym{ttm}{TTM}{time to market}
\newacronym{ttn}{TTN}{The Things Network}
\newacronym{tx}{TX}{transmitter}
\newacronym{uav}{UAV}{unmanned aerial vehicle}
\newacronym{uc}{UC}{use case}
\newacronym{ucie}{UCIe}{Universal Chiplet Interconnect Express}
\newacronym{udp}{UDP}{User Datagram Protocol}
\newacronym{ue}{UE}{user equipment}
\newacronym{ugv}{UGV}{unmanned ground vehicle}
\newacronym{uhd}{UHD}{USRP hardware driver}
\newacronym{uhf}{UHF}{ultra-high frequency}
\newacronym{ul}{UL}{uplink}
\newacronym{ula}{ULA}{uniform linear array}
\newacronym{uMIMO}{$\mu$-MIMO}{ultra-massive Multiple-Input Multiple-Output}
\newacronym{ummtc}{umMTC}{ultra massive machine type communication}
\newacronym{un}{UN}{United Nations}
\newacronym{upa}{UPA}{uniform planar array}
\newacronym{ura}{URA}{uniform rectangular array}
\newacronym{urllc}{URLLC}{ultra-reliable low-latency communications}
\newacronym{usrp}{USRP}{universal software radio peripheral}
\newacronym{uv}{UV}{unmanned vehicle}
\newacronym{uwb}{UWB}{ultrawideband}
\newacronym{v2v}{V2V}{vehicle-to-vehicle}
\newacronym{vep}{VEP}{virtual edge platform}
\newacronym{vlc}{VLC}{visible light communication}
\newacronym{vlp}{VLP}{visible light positioning}
\newacronym{vna}{VNA}{vector network analyzer}
\newacronym{votable}{VOTable}{Virtual Observatory Table}
\newacronym{vr}{VR}{virtual reality}
\newacronym{wb}{WB}{wideband}
\newacronym{wban}{WBAN}{wireless body area network}
\newacronym{wimax}{WiMAX}{Worldwide Interoperability for Microwave Access}
\newacronym{wlan}{WLAN}{wireless LAN}
\newacronym{wpt}{WPT}{wireless power transfer}
\newacronym{wr}{WR}{White Rabbit}
\newacronym{wrsn}{WRSN}{wireless rechargeable sensor network}
\newacronym{wsn}{WSN}{wireless sensor network}
\newacronym{xets}{XETS}{cross exponentially tapered slot}
\newacronym{xr}{XR}{extended reality}
\newacronym{z3ro}{Z3RO}{zero third-order distortion}
\newacronym{zf}{ZF}{zero-forcing}
\newacronym{zmcscg}{ZMCSCG}{zero mean circularly symmetric complex Gaussian}

\usepackage{siunitx}
\DeclareSIUnit{\decibelm}{dBm}
\DeclareSIUnit{\decibeli}{dBi}
\DeclareSIUnit{\farad}{F}
\DeclareSIUnit{\milliamperehour}{mAh}
\DeclareSIUnit{\gCOeq}{gCO\textsubscript{2}eq}
\DeclareSIUnit{\kgCOeq}{kgCO\textsubscript{2}eq}

\usepackage{tikz, pgfplots}
\usetikzlibrary{backgrounds}
\usetikzlibrary{calc}

\usepackage{xspace}
\ifarxiv
\newcommand{\arxiv}[1]{ #1\xspace}
\newcommand{\rmcite}[1]{~\cite{#1}}
\usepackage[backend=biber, style=ieee, citestyle=numeric-comp, isbn=true, doi=true,citecounter=true,maxnames=3]{biblatex}
\else 
\newcommand{\arxiv}[1]{}
\newcommand{\rmcite}[1]{}
\usepackage[backend=biber, style=ieee, maxnames=1, citestyle=numeric-comp, isbn=false, doi=false,citecounter=true]{biblatex}
\fi

\AtBeginBibliography{\footnotesize}

\usepackage{threeparttable}
\usepackage{booktabs}
\usepackage{multicol}
\usepackage{multirow}

\usepackage{pgf-pie, etoolbox}
\usepackage{pgf}

\usepackage[capitalize]{cleveref}

\usepackage{placeins}

\begin{document}

\title{IoT on the Road to Sustainability:\\Vehicle or Bandit?}

\author{
    \IEEEauthorblockN{Jona Cappelle$^*$, Liesbet Van der Perre$^*$, Emma Fitzgerald$^{\dagger}$, Simon Ravyts$^*$, Weronika Gajda$^*$, Valentijn De Smedt$^*$, Bert Cox$^*$, Gilles Callebaut$^*$}\\
    \IEEEauthorblockA{
        $^*$ \textit{KU Leuven, ESAT, Belgium},  name.surname@kuleuven.be
    }\\
    \IEEEauthorblockA{
        $^{\dagger}$ \textit{EIT, Lund University, Box 118, 221 00 Lund, Sweden}, emma.fitzgerald@eit.lth.se
    }
    \arxiv{\thanks{This manuscript was submitted to IEEE IoT Magazine on 31st of May 2024.}}
}

\maketitle

\begin{abstract} The \gls{iot} can support the evolution towards a digital and green future. %
However, the introduction of the technology clearly has in itself a direct adverse ecological impact. This paper assesses this impact at both the \gls{iot}-node and at the network side. For the nodes, we show that the electronics production of devices comes with a carbon footprint that can be much higher than during operation phase. We highlight that the inclusion of IoT support in existing cellular networks comes with a significant ecological penalty, raising overall energy consumption by more than 15\%. %
These results call for novel design approaches for the nodes and for early consideration of the support for \gls{iot} in future networks. Raising the `Vehicle or bandit?' question on the nature of \gls{iot} in the broader sense of sustainability, we illustrate the need for multidisciplinary cooperation to steer applications in desirable directions. %

\end{abstract}


\begin{figure*}[b]
    \centering
    \includegraphics[width=0.88\textwidth, trim={1cm 4cm 2cm 3cm}, clip]{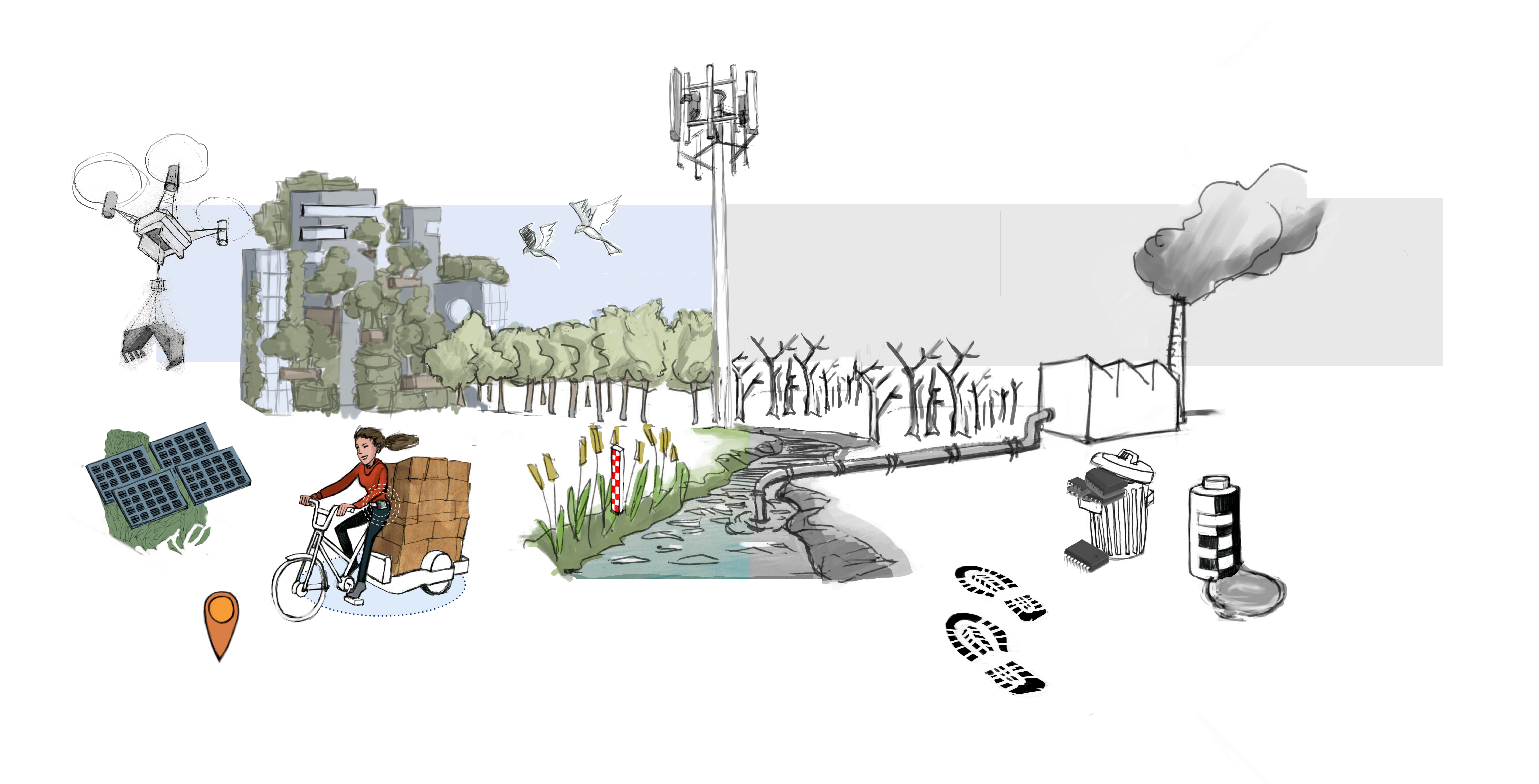}
    \caption{Overview: IoT vehicle or bandit?}%
    \label{fig:roadtotake}
\end{figure*}  \FloatBarrier

\section{Introduction}
\glsresetall

The \Gls{iot}, enabling remote monitoring~\cite{RemoteIoT_art} and efficiency improvements, has been promoted as a vehicle\rmcite{Salam2020Forestry} towards the defined 17 \glspl{sdg} defined by the \gls{un} as a global call to action~\cite{SDGs}. So far, however, the technology often plays the role of a bandit hijacking us on this journey. %

This paper assesses whether the direct ecological penalty of \gls{iot} technology can be outweighed by the impact on other systems' footprints. Contradictory goals exist when connecting \gls{iot} devices to address sustainability. For example, enabling node-initiated communication conflicts with sleep opportunities in the network. Also, many eco-digital innovations address the symptoms of a problem, for example finding free parking spots, rather than \textit{the} problem - too many cars in cities.  %
At the eve of a potential massive deployment of the technology, it is crucial to be aware of possible directions, as illustrated in Fig.~\ref{fig:roadtotake}: can we minimize the bandit-nature and envision \gls{iot} solutions as a vehicle to address fundamental problems? The main contribution of this paper is threefold: (i) it analyses the direct full-life-cycle impact of \gls{iot} nodes, discusses a modular design approach, and demonstrates the need 
to quantitatively assess the net balance in applications, (ii) it studies the consequences of introducing \gls{iot} networks%
 considering deployments in licensed and unlicensed bands, and (iii) it identifies opportunities for and risks to all SDGs in the profit, planet, and people categories. 

In this paper, we first assess the direct impact of the technology at the \gls{iot} node and in networks. Next, we discuss the potential indirect impact of \gls{iot} on the broad set of \glspl{sdg}. %
Finally, we summarize the conclusions, and formulate suggestions for future R\&D.%

\section{IoT Node Direct Ecological Impact}
\label{sec:node}

\begin{figure*}
\centering

\includegraphics[scale=0.75]{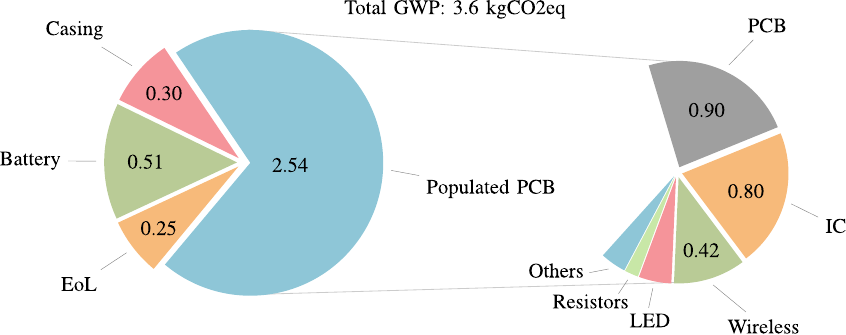}

\caption{GWP Typical IoT Node (left) with detailed populated PCB analysis (right) in \si{\kgCOeq}.} %
\label{fig:piechartgwp}
\end{figure*}

Impressive progress on \glspl{ic}, miniature sensors and actuators enabled by \gls{mems}-technology, has created endless possibilities to design \gls{iot}-nodes with a small form factor and at low cost. However, the ecological footprint of electronic devices is considerably larger than their physical size would suggest~\cite{Pirson2023Evaluating}. 
Current \Gls{iot}-node architectures typically aim for very high hardware integration and devices are often fully sealed, which comes with advantages in size and physical protection of the devices. However, this approach prohibits their eventual ecological dismantling, inhibits reusability, and does not address recyclability. We present a \gls{lca} to critically review the ecological impact of all three stages in the life cycle of an \gls{iot} device~\cite{carbon-footprint-iot}: design and production, operation, and \gls{eol}. 
We further present potentially more sustainable design paradigms, operation, and \gls{eol} approaches.

\subsection{Design and production phase} %
We performed an \gls{lca} on a representative \gls{iot} device, following the methodology of ISO 14040 and 14044 standards. The \gls{gwp} was modeled with the Sphera professional and Electronics database~\cite{sphera} using the ReCiPe 2016 midpoint method.
The results as depicted in~\cref{fig:piechartgwp}, not considering the installation, show that the production of electronics, in particular \glspl{ic} and \glspl{pcb}, comes with a high \gls{gwp}. Studies~\cite{carbon-footprint-iot} based on current usage models have indicated that production is typically responsible for \SIrange{60}{70}{\percent} of the carbon footprint of mobile devices.
The carbon footprint for low to medium-complexity devices, for example a water tank level meter, %
 with few sensors and basic wireless connectivity is estimated to account for \SIrange{1}{10}{\kgCOeq}~\cite{carbon-footprint-iot}. In terms of water usage, a typical microcontroller uses \SI{23}{\liter} during its life cycle, of which the majority is during production~\cite{sphera}. 

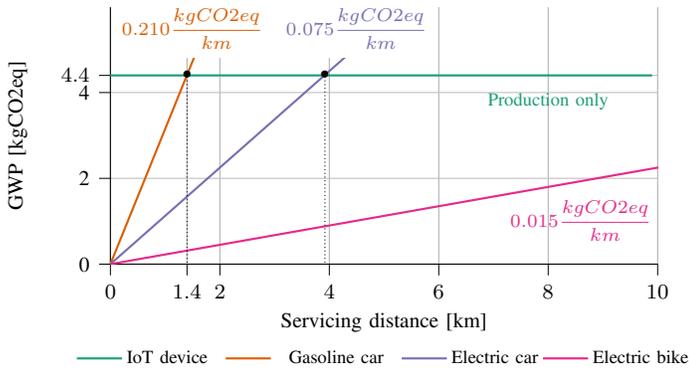
\begin{figure}[t]
    \centering

\usetikzlibrary{positioning}

       \begin{tikzpicture}[X/.style = {circle, fill=black, inner sep=1.5pt, 
        label={[font=\footnotesize]above right:#1},
        node contents={}},
        trim axis left, trim axis right]
    
        \ifx\usegreyscale\undefined
        \definecolor{color0}{HTML}{1b9e77}
        \definecolor{color1}{HTML}{d95f02}
        \definecolor{color2}{HTML}{7570b3}
        \definecolor{color3}{HTML}{e7298a}
        \definecolor{color4}{HTML}{66a61e}
        \definecolor{color5}{HTML}{e6ab02}
        \else
        \definecolor{color0}{HTML}{525252}
        \definecolor{color1}{HTML}{d9d9d9}
        \definecolor{color2}{HTML}{bdbdbd}
        \definecolor{color3}{HTML}{969696}
        \definecolor{color4}{HTML}{636363}
        \definecolor{color5}{HTML}{252525}
        \fi
    
        \begin{axis}[
            use as bounding box,
            axis line style={black},
            axis lines* = {left},
            legend cell align={left},
            legend style={ at={(0.5,-0.3)},anchor=north, fill opacity=0.8, draw opacity=1, text opacity=1, draw=white!80.00000!black, /tikz/column 2/.style={column sep=5pt}, /tikz/column 3/.style={column sep=5pt},/tikz/column 4/.style={column sep=5pt}},
            legend style={draw=none, nodes={scale=0.7, transform shape}},
            legend columns=4, 
            tick align=outside,
            tick pos=left,
            width = \columnwidth,
            height=5cm,
            x grid style={white!69.01960784313725!black},
            grid,
            xtick style={color=black},
            ylabel={GWP [kgCO2eq]},
            xlabel={Servicing distance [km]},
            ytick style={color=black},
            label style={font=\footnotesize},
            tick label style={font=\footnotesize},
            extra x ticks={1.4},
            extra y ticks={4.4},
            xmin=0,
            xmax=10,
            ymin=0,
            ymax=5.99,
            yticklabels={0,0,2,4}
            ]

    \addplot[color0, thick] coordinates
    {
        (0.0,4.4)
        (9.8989898989899,4.4)
    };
            
    \addlegendentry{IoT device}
    
    \addplot[color1, thick] coordinates
    {
        (0.0,0.0)
        (20.0,63.0)
    };
    \addlegendentry{Gasoline car}
    
    \addplot[color2, thick] coordinates
    {
        (0.0,0.0)
        (20.0,22.5)
    };
    \addlegendentry{Electric car}

    \addplot[color3, thick] coordinates
    {
        (0.0,0.0)
        (20.0,4.5)
    };
    \addlegendentry{Electric bike}

        \coordinate (loc_cross_1) at (axis cs:1.4,4.37);
        \coordinate (loc_cross_2) at (axis cs:3.92,4.37);

        \coordinate (loc_cross_1_text) at (axis cs:1.5,3);
        \coordinate (loc_cross_2_text) at (axis cs:4.5,3);

        \node at (loc_cross_1) {\textbullet};

        \node at (loc_cross_2) {\textbullet};

        \draw [densely dotted] (loc_cross_1) -- (axis cs:1.4,0);
        \draw [densely dotted] (loc_cross_2) -- (axis cs:3.92,0);

    \node[align=left, anchor=south,fill=white!80] at (axis cs:1.5,4.8) {\color{color1} \scriptsize $0.210\dfrac{kgCO2eq}{km}$};

    \node[align=left, anchor=south,fill=white!80] at (axis cs:4.5,4.8) {\color{color2} \scriptsize $0.075\dfrac{kgCO2eq}{km}$};

    \node[align=left, anchor=north] at (axis cs:8.6,1.7) {\color{color3} \scriptsize $0.015\dfrac{kgCO2eq}{km}$};

    \node[align=left, anchor=north] at (axis cs:8.0,4.2) {\color{color0} \scriptsize Production only};
            
    \end{axis}

    \end{tikzpicture}

    \caption{Low power IoT case. Comparison of the environmental impact of smart meter system with a yearly manual intervention.}
    \label{fig:comparison_cfp}
\end{figure}

\textbf{First question: to \gls{iot} or not to \gls{iot}?}
Considering the ecological footprint associated with the production of the devices, a first question to answer when considering an \gls{iot} application is whether the indirect ecological benefit outweighs the `cradle to gate' impact. We provide the results of an analysis for a typical smart meter use case in~\cref{fig:comparison_cfp}. It compares the introduction of an \gls{iot}-based monitoring with yearly manual interventions by car over 15~years, either driving on gasoline or electric, or by an electric bike. The results show that driving a gasoline car is only ecologically justifiable for distances below \SI{1.4}{\kilo\meter}, a distance one can argue that could be covered on foot. Up to \SI{4}{\kilo\meter} an electric car could be an adequate choice, yet is significantly outperformed by the electric bike for distances up to \SI{18}{\kilo\meter}.\footnote{For short distances the motor is not needed, the biker may have a personal calorie surplus to spend.} For remote smart meters, the ecological penalty of driving any car is getting substantially worse than the impact of the \gls{iot} solution. To conclude, \gls{iot}-based smart meters will only bring benefits in \gls{gwp} in densely populated areas if more regular updates or additional features are required, such as detection of water leaks.
We note that while the carbon footprint numbers are not high in absolute terms for an individual device, there is a risk that the overall impact will escalate as the deployment of devices reaches large scales, especially when they incorporate larger processing power and memory using smaller semiconductor technology nodes. These smaller technologies imply more complex manufacturing, resulting in a larger carbon footprint~\cite{carbon-footprint-iot}.

\textbf{Second question: can a modular IoT node design bring ecological benefits?}
 A `fire and forget' approach is applied in many \gls{iot} applications, and the risk of trash left behind is far larger than in, e.g., smartphones, where there are incentives to resell or reuse. It is imperative to incorporate \gls{eol} considerations from the initial design stages.  
We study a more modular \gls{iot} architecture as an enabler to reduce the carbon footprint and e-waste.
The aim is to make it possible to upgrade, replace, or recycle parts when broken or obsolete.
Note, however, that modular design is not guaranteed to be the most sustainable design option, since it inevitably comes with overhead in material consumption. The latter can pay off through a significantly longer use of devices and modules, and/or easier dismantling and recycling.

We assess the environmental impact of \gls{iot} node energy provisioning by an electric car, a traditional solar panel, and UAV-based interventions. Considering practical feasibility for automation, next to a wired connection between the battery and the \gls{iot} node, a wireless interconnection via inductive \gls{wpt} is included in the UAV case.
We calculated an overhead of $\pm$ \SI{700}{\gCOeq} for the wireless interconnection, equivalent to a $\pm$ \SI{10}{km} drive with a small electric car. The results, depicted in~\cref{tab:comparison_cfp},
 indicate that utilizing solar panels is the most environmentally sustainable choice whenever feasible, since little to no servicing has to be performed. The electric car is shown to be the least sustainable option. %
For applications that can not rely on energy harvesting, UAV-based servicing can be a viable alternative. Due to the inefficiency of the \gls{wpt} and additional hardware needs, the wirelessly connected UAV-based approach has a higher environmental impact.

\subsection{During operation}
 Batteries are often the limiting factor for operation time of  \gls{iot} nodes. 
Therefore, minimizing energy consumption is an effective way to either extend their operational lifespan or reduce the carbon footprint by the ability to use smaller batteries.
Since data transmission is a major contributor to the total energy consumption, energy-efficient wireless communication is key to enabling sustainable growth of \gls{iot} deployments~\cite{RemoteIoT_art}.
Energy-neutral nodes that rely on harvested energy, due to the elimination of batteries, can provide environmental benefits. %
This reasoning, while having its merit from a specific viewpoint, is based on over-simplifications and highly depends on the energy provisioning technology used. 
Very efficient perovskite solar cells\cite{perovskite_solar_cells},
e.g., have a high potential in both increasing the efficiency closer to the Shockley–Queisser limit and simplifying the manufacturing process compared to traditional crystalline Si-based solar cells. %
\Gls{rf} energy harvesting can deliver only small amounts of energy at relatively short ranges.
Careful consideration is crucial. Using mixed-grid energy in Europe without factoring in extra power circuitry, we determined %
that the wireless power transfer efficiency needs to exceed \SI{0.5}{\percent} to achieve a lower carbon footprint per delivered watt-hour compared to using non-rechargeable batteries~\cite{sphera}.

\subsection{End-of-life}
There is little incentive, from a practical or economic perspective, to care about \gls{iot} nodes when they are broken or no longer needed. 
\gls{iot} systems are typically optimized for ease of installation and operation, pursuing plug-and-play solutions.
In locations where human intervention is cumbersome, devices can be hard to recover. Collecting and recycling precious materials is not a systematic practice. The `fire and forget' credo is too forgetful, resulting in e-waste. 
Since 2014, the global %
generation of e-waste has grown by 9.2 million metric tons (Mt) (21\%). The fate of over four-fifths (82.6\% or 44.3 Mt) of e-waste generated in 2019 is unknown, as is its impact on the environment~\cite{ITU_eWaste}. A massive deployment of \gls{iot} systems bears a risk of many nodes left behind, with potentially toxic components such as the batteries~\cite{liion_bad}. Even small non-rechargeable batteries, having a limited carbon footprint (\cref{fig:piechartgwp}), contain a lot of rare and toxic materials.

Biodegradables have been studied for two decades~\cite{advances_in_biodegradables, myceliotronics}, but a performance gap still exists between the state-of-the-art Si-electronics compared to their biodegradable equivalents. 
A promising field of biodegradable electronics is energy harvesting and sensors. While the efficiency of biodegradable \gls{pv} materials is low compared to mainstream devices, due to the reduced waste and cost, they can compete with mainstream solutions~\cite{advances_in_biodegradables}. 
Biodegradables electronics, e.g., myceliotronics with materials made out of fungus, could be a key enabler for the \gls{eol} process, as they could be left in nature. %
However, even if the biodegradable alternative has a reduced impact during production, the shorter lifespan, and thus the need for more frequent replacements, could ultimately surpass the carbon footprint of a non-biodegradable approach. 

To conclude the assessment of the direct ecological impact of \gls{iot} nodes, it is clear that their footprint may be many times larger than only their energy consumption during operation: the full life cycle needs to be considered, which urges novel design paradigms and \gls{eol} solutions.

\begin{table}[t!]
\caption{Intense usage IoT case (500J/day), comparison between servicing methods R: rechargeable, NR: non-rechargeable, fl: flight, $\ast$: wireless, $\triangleright$: contacts. In [kgCO2eq], unless noted otherwise. The 1-time \gls{gwp} for installation is not taken into account.}%
\resizebox{\columnwidth}{!}{%
\begin{tabular}{@{}lcccc@{}}
\toprule
                    & \textbf{E-car}         & \textbf{Solar panel} &  \textbf{UAV} ($\ast$)  &  \textbf{UAV} ($\triangleright$) \\ \midrule
Overhead            & 0                      & 1/Wp + 0.2        & 0.7                    & 0                      \\
Battery             & \begin{tabular}[c]{@{}c@{}} 0.8 \\ (22.5Wh NR) \end{tabular}  & \begin{tabular}[c]{@{}c@{}} 0.1 \\ (1Wh R) \end{tabular}         & \begin{tabular}[c]{@{}c@{}} 1 \\ (10Wh R) \end{tabular}           & \begin{tabular}[c]{@{}c@{}} 0.8 \\ (22.5Wh NR) \end{tabular}         \\
Service             & 0.075/km                 & 0                    & 0.060/h                   & 0.060/h                   \\ \midrule
\begin{tabular}[l]{@{}c@{}} Lifetime\\ (battery) \end{tabular} & 1.3 years              & 15+ years            & 2.4 months              & 1.3 years               \\
Service (1km)  & 0.8625                 & 0                    & 0.75 (10min)            & 0.115(10min)            \\
Service (5km)  & 4.3125                 & 0                    & 3.75 (50min)            & 0.575 (50min)           \\
Total (1km)         & 10.06                  & 1.3                  & 2.45                    & 9.315                   \\
Total (5km)         & 13.51                  & 1.3                  & 5.45                    & 9.775                   \\ \bottomrule
\end{tabular}
}
\label{tab:comparison_cfp}
\end{table}

\definecolor{nrsColor}{HTML}{8ec6d7}
\definecolor{nsssColor}{HTML}{9f9f9f}
\definecolor{npbchColor}{HTML}{f5949a}
\definecolor{noSigColor}{HTML}{b9cb96}
\definecolor{npssColor}{HTML}{f7ba7a}

\begin{figure*}[t]
\centering
     \begin{subfigure}[b]{0.3\textwidth}
         \centering
         \begin{tikzpicture}

\def\NRS{22.14}
\def\NPBCH{6.43}
\def\NPSS{6.43}
\def\NSSS{3.2}
\def\free{61.8}

\tikzset{
     lines/.style={draw=none},
}

\tikzstyle{every node}=[font=\scriptsize]

 \pie [/tikz/every pin/.style={align=center},
  every only number node/.style={text=white},
  text=pin,%
  rotate=300,
  explode=0.05,
  radius=1.6,
  color={nrsColor,npbchColor, npssColor, nsssColor, noSigColor},
  style={lines}]
    {\NRS/NRS,
     \NPBCH/NPBCH, 
     \NPSS/NPSS,
     \NSSS/NSSS,
     \free/No\\signalling}

\end{tikzpicture}
         \vspace{-0.75cm}
         \caption{Share of unoccupied \textit{free} time slots versus the different downlink signals.}
     \end{subfigure}\hfill%
     \begin{subfigure}[b]{0.6\textwidth}
         \centering
          \begin{tikzpicture}

\definecolor{nrsColor}{HTML}{8ec6d7}
\definecolor{piegreen1}{HTML}{c8e7a7}
\definecolor{npbchColor}{HTML}{f5949a}
\definecolor{noSigColor}{HTML}{b9cb96}
\definecolor{npssColor}{HTML}{f7ba7a}

\def\numRows{1}
\def\numCols{280}
\pgfmathsetmacro{\blockSize}{scalar(0.8\textwidth/280cm)}
\def\blockHeight{1.5}

\foreach \col in {0,...,279} {
     \fill[noSigColor] (\col*\blockSize, 0) rectangle ++(\blockSize, \blockHeight/2);
  }

  \foreach \col in {3,...,13} {
     \fill[npbchColor] (\col*\blockSize, 0) rectangle ++(\blockSize, \blockHeight);
  }

  \foreach \col in {19,...,20} {
     \fill[nrsColor] (\col*\blockSize, 0) rectangle ++(\blockSize, \blockHeight);
  }

  \foreach \col in {26,...,27} {
     \fill[nrsColor] (\col*\blockSize, 0) rectangle ++(\blockSize, \blockHeight);
  }

  \foreach \col in {33,...,34} {
     \fill[nrsColor] (\col*\blockSize, 0) rectangle ++(\blockSize, \blockHeight);
  }

  \foreach \col in {40,...,41} {
     \fill[nrsColor] (\col*\blockSize, 0) rectangle ++(\blockSize, \blockHeight);
  }

  \foreach \col in {47,...,48} {
     \fill[nrsColor] (\col*\blockSize, 0) rectangle ++(\blockSize, \blockHeight);
  }

  \foreach \col in {54,...,55} {
     \fill[nrsColor] (\col*\blockSize, 0) rectangle ++(\blockSize, \blockHeight);
  }

  \foreach \col in {61,...,62} {
     \fill[nrsColor] (\col*\blockSize, 0) rectangle ++(\blockSize, \blockHeight);
  }

  \foreach \col in {68,...,69} {
     \fill[nrsColor] (\col*\blockSize, 0) rectangle ++(\blockSize, \blockHeight);
  }

  \foreach \col in {73,...,83} {
     \fill[npssColor] (\col*\blockSize, 0) rectangle ++(\blockSize, \blockHeight);
  }

  \foreach \col in {89,...,90} {
     \fill[nrsColor] (\col*\blockSize, 0) rectangle ++(\blockSize, \blockHeight);
  }

  \foreach \col in {96,...,97} {
     \fill[nrsColor] (\col*\blockSize, 0) rectangle ++(\blockSize, \blockHeight);
  }

  \foreach \col in {103,...,104} {
     \fill[nrsColor] (\col*\blockSize, 0) rectangle ++(\blockSize, \blockHeight);
  }

  \foreach \col in {110,...,111} {
     \fill[nrsColor] (\col*\blockSize, 0) rectangle ++(\blockSize, \blockHeight);
  }

  \foreach \col in {117,...,118} {
     \fill[nrsColor] (\col*\blockSize, 0) rectangle ++(\blockSize, \blockHeight);
  }

  \foreach \col in {124,...,125} {
     \fill[nrsColor] (\col*\blockSize, 0) rectangle ++(\blockSize, \blockHeight);
  }

  \foreach \col in {129,...,139} {
     \fill[nsssColor] (\col*\blockSize, 0) rectangle ++(\blockSize, \blockHeight);
  }
  \foreach \col in {143,...,153} {
     \fill[npbchColor] (\col*\blockSize, 0) rectangle ++(\blockSize, \blockHeight);
  }

  \foreach \col in {159,...,160} {
     \fill[nrsColor] (\col*\blockSize, 0) rectangle ++(\blockSize, \blockHeight);
  }

  \foreach \col in {166,...,167} {
     \fill[nrsColor] (\col*\blockSize, 0) rectangle ++(\blockSize, \blockHeight);
  }

  \foreach \col in {173,...,174} {
     \fill[nrsColor] (\col*\blockSize, 0) rectangle ++(\blockSize, \blockHeight);
  }

  \foreach \col in {180,...,181} {
     \fill[nrsColor] (\col*\blockSize, 0) rectangle ++(\blockSize, \blockHeight);
  }

  \foreach \col in {187,...,188} {
     \fill[nrsColor] (\col*\blockSize, 0) rectangle ++(\blockSize, \blockHeight);
  }

  \foreach \col in {194,...,195} {
     \fill[nrsColor] (\col*\blockSize, 0) rectangle ++(\blockSize, \blockHeight);
  }

  \foreach \col in {201,...,202} {
     \fill[nrsColor] (\col*\blockSize, 0) rectangle ++(\blockSize, \blockHeight);
  }
  \foreach \col in {208,...,209} {
     \fill[nrsColor] (\col*\blockSize, 0) rectangle ++(\blockSize, \blockHeight);
  }
  \foreach \col in {213,...,223} {
     \fill[npssColor] (\col*\blockSize, 0) rectangle ++(\blockSize, \blockHeight);
  }
  \foreach \col in {229,...,230} {
     \fill[nrsColor] (\col*\blockSize, 0) rectangle ++(\blockSize, \blockHeight);
  }

  \foreach \col in {236,...,237} {
     \fill[nrsColor] (\col*\blockSize, 0) rectangle ++(\blockSize, \blockHeight);
  }

  \foreach \col in {243,...,244} {
     \fill[nrsColor] (\col*\blockSize, 0) rectangle ++(\blockSize, \blockHeight);
  }

  \foreach \col in {250,...,251} {
     \fill[nrsColor] (\col*\blockSize, 0) rectangle ++(\blockSize, \blockHeight);
  }

  \foreach \col in {257,...,258} {
     \fill[nrsColor] (\col*\blockSize, 0) rectangle ++(\blockSize, \blockHeight);
  }

  \foreach \col in {264,...,265} {
     \fill[nrsColor] (\col*\blockSize, 0) rectangle ++(\blockSize, \blockHeight);
  }

  \foreach \col in {271,...,272} {
     \fill[nrsColor] (\col*\blockSize, 0) rectangle ++(\blockSize, \blockHeight);
  }
  \foreach \col in {278,...,279} {
     \fill[nrsColor] (\col*\blockSize, 0) rectangle ++(\blockSize, \blockHeight);
  }

\foreach \col in {0,...,19} {
 \draw[black!40] (14*\col*\blockSize, 0) rectangle ++(14*\blockSize, \blockHeight);
}

\foreach \col [count=\xi from 0] in {7,21,...,133} {
    \node  at (\col*\blockSize,-0.2) {\scriptsize \xi};
    \ifnum\xi=0
    \node[rotate=90, text=black]  at (\col*\blockSize,\blockHeight/2) {\scriptsize NPBCH}; 
    \fi
    \ifnum\xi=5
    \node[rotate=90, text=black]  at (\col*\blockSize,\blockHeight/2) {\scriptsize NPSS}; 
    \fi
    \ifnum\xi=9
    \node[rotate=90, text=black]  at (\col*\blockSize,\blockHeight/2) {\scriptsize NSSS}; 
    \fi
}

\foreach \col [count=\xi from 0] in {147,161,...,273} {
    \node at (\col*\blockSize,-0.2) {\scriptsize \xi};
     \ifnum\xi=0
    \node[rotate=90, text=black]  at (\col*\blockSize,\blockHeight/2) {\scriptsize NPBCH}; 
    \fi
    \ifnum\xi=5
    \node[rotate=90, text=black]  at (\col*\blockSize,\blockHeight/2) {\scriptsize NPSS}; 
    \fi
}

\node[right] at (\numCols*\blockSize,-0.2) {\scriptsize Subframe \#};

\draw[->] (0,-0.5) -- (\numCols*\blockSize,-0.5) node[right] {\scriptsize Time};

\end{tikzpicture}
          \vspace{0.4cm} %
          \vspace{-0.75cm}
          \caption{Time representation of the frame structure of \gls{nbiot} downlink signalling.}
     \end{subfigure}
\caption{Illustration of time resources occupied by downlink signals of \gls{nbiot}. Almost \SI{40}{\percent} of the time is spent in downlink signalling events, prohibiting the base station to go to sleep during these periods. See \cref{sec:nbiot} for details regarding the different signals. \arxiv{(This example is for the guard band and stand-alone deployment of \gls{nbiot}.)}}%
\label{fig:piechartNBIoT}
\end{figure*}
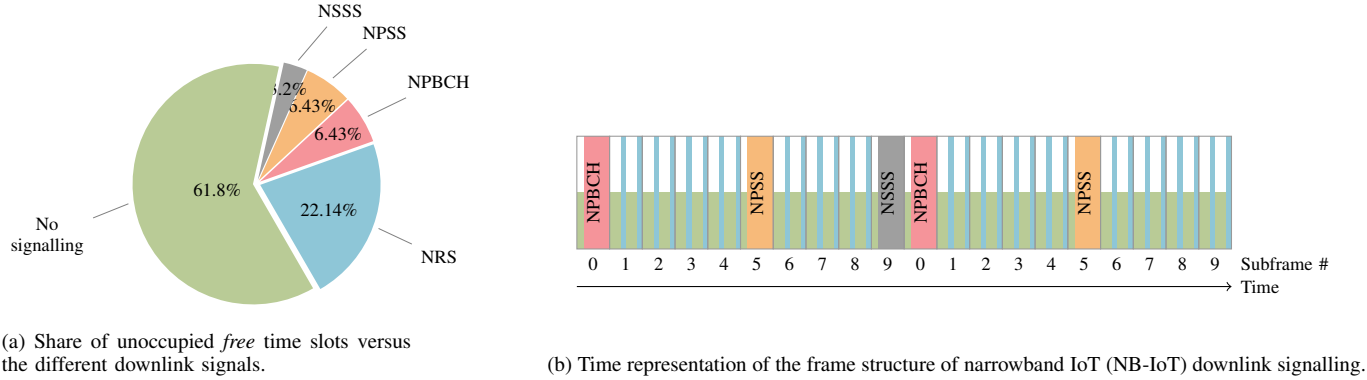

\section{Network-Side Direct Impact}
\label{sec:network}
Wireless connectivity is the magic ingredient that enables \gls{iot} applications in many environments. 
\Gls{lpwan} technologies offer the sought-after combination of low-power and long-range connectivity for applications generating (primarily) sporadic uplink data. 
\Gls{lpwan} networks are established either as dedicated infrastructures operating in unlicensed bands, or embedded in existing cellular networks. It is clear that installing new networking equipment has a significant ecological footprint. Also, while much less obvious, the introduction of support for \gls{iot} in existing cellular networks comes with a substantial ecological impact, as we demonstrate below.

\subsection{Network Deployment: Top-Down versus Bottom-Up}
Several \gls{lpwan} technologies, operating in the unlicensed \SI{868}{MHz} band, have been proposed. These require dedicated gateway installations, integrating electronics with an ecological impact~\cite{9979766}. 
While not intended for commercial network deployments, many operators have rolled out networks in these unlicensed bands. Most notably, cellular operators deploy LoRaWAN and Sigfox technologies~\cite{RemoteIoT_art}, using conventional top-down approaches, i.e., thorough network planning.

Crowdsourced \gls{lpwan} deployments have gained momentum offering an alternative to operator-based networks. The license-free band promotes open access. For example, in remote and/or less prosperous areas where commercial deployments may lag behind, they can lower the entry barrier  %
to set up \gls{iot}-enabled applications. This has been successfully demonstrated by initiatives such as \gls{ttn}, in which tens of thousands of gateways are connected, and, e.g., smart irrigation is supported. %
An organically growing crowdsourced network may lead to a suboptimal usage of resources as no top-down network planning is performed. To illustrate this, there are, at the moment of writing, \num{79097} gateways registered at the \gls{ttn}. Only \num{16340} of these are online and in operation. This entails that \SI{79}{\percent} of the gateways are offline, despite being produced and registered to the \gls{ttn}, each having a \SI{7}{\kgCOeq} footprint for an average outdoor gateway~\cite{chiew2021life}. %
Despite this downside of bottom-up deployments, the organically growing crowdsourced networks may lead to `just enough' coverage and infrastructure installation. %

\subsection{How low-power technologies yield high-power networks}\label{sec:nbiot}

An important advantage of NB-IoT networks, over non-cellular technologies, is that they can provide quality-of-service guarantees. This may be vital for applications, for example in the e-health domain. 

NB-IoT is an \gls{lpwan} technology that can in principle be implemented as a software upgrade on existing cellular network infrastructures. This has the clear benefit that no new hardware, with associated ecological impact. However, the assessment of the energy penalty of introducing NB-IoT in 4G networks demonstrates that the integration thereof directly raises their baseline energy consumption considerably. To lower the overall power consumption of cellular networks, strategies were devised to allow \glspl{bs} to enter a low-power sleep state.
First, non-time-critical (downlink) data, generating bursty and sporadic traffic, is buffered rather than spreading it over time. This results in a longer time in which the \gls{bs} can be in the energy-saving idle mode, i.e., when there are no devices to be served. This is exploited by the sleep strategy, where the power amplifiers are turned off when no data or control signals need to be sent.
However, due to NB-IoT signaling, the \glspl{bs} are prohibited from going into sleep, increasing power consumption by \SI{16.8}{\percent} in idle mode. This is illustrated in~\cref{fig:piechartNBIoT}. The \gls{bs} is occupied almost \SI{40}{\percent} of the time by signaling to allow the \gls{ue} --- the \gls{iot} device --- to find, synchronize and connect to the network, i.e., \gls{nrs}, \gls{npbch}, \gls{nsss} and \gls{npss}.

This penalty can not be neglected, neither in relative nor absolute terms, as ICT networks are estimated to consume \SI{1.15}{\percent} of the global grid electricity supply today and they have been reported as fast growing.  %

\section{Indirect Impact: Vehicle or Bandit?}
\label{sec:inducedImpact}

\begin{figure*}[hbt]
    \centering
    \begin{tikzpicture}%
    
    \tikzstyle{line_width_bars}=[line width=1.25mm]
    \tikzstyle{lines}=[black!25, line width=0.5mm]
    \tikzstyle{bar_vehicle}=[Green!25, line_width_bars]
    \tikzstyle{bar_vehicle_future}=[Green!50, line_width_bars]
    \tikzstyle{bar_bandit}=[Red!25, line_width_bars]
    \tikzstyle{bar_bandit_future}=[Red!50, line_width_bars]
    
    \def\spacebetweenlines{0.2} %
    \def\extralegendspacing{0.1} %
    \def\offset{0.3} %
    \def\offsetmiddletext{0.3}
    \def\spacebetweensubsections{0.8}
    \def\textboxwidth{5cm} %
    \def\centertextboxwidth{5.6cm}
    \def\placefrommargin{85}
    \newcommand{\entry}[9]{%
        \node[fill=white] at (0,-#1+0.4-\offsetmiddletext+\offset) {\textbf{#6}};
        \draw[bar_vehicle] (0,-#1+\offset-0.8) -- (#2,-#1+\offset-0.8);
        \draw[bar_vehicle_future] (0,-#1-\spacebetweenlines+\offset-0.8) -- (#3,-#1-\spacebetweenlines+\offset-0.8);
        \draw[bar_bandit] (-#4,-#1+\offset-0.8) -- (0,-#1+\offset-0.8);
        \draw[bar_bandit_future] (-#5,-#1-\spacebetweenlines+\offset-0.8) -- (0,-#1-\spacebetweenlines+\offset-0.8);
        \draw[lines, color=black!5] (-\textwidth/2,-#1-1.3+\offset) -- (\textwidth/2,-#1-1.3+\offset);
        \node[anchor=center, align=left, text width=\textboxwidth, execute at begin node=\setlength{\baselineskip}{1ex}] at (-\textwidth/2+\placefrommargin,-#1+\offset-0.3) {\footnotesize{#9}};
        \node[anchor=center, align=right, text width=\textboxwidth, execute at begin node=\setlength{\baselineskip}{1ex}] at (\textwidth/2-\placefrommargin,-#1+\offset-0.3) {\footnotesize{#8}};
        \node[fill=white, anchor=north, text width=\centertextboxwidth, execute at begin node=\setlength{\baselineskip}{1ex}, align=center] at (0,-#1+\offset+0.2-\offsetmiddletext) {\footnotesize{#7}};
    }
    \newcommand{\legend}[4]{%
        \node[anchor=west] at (#3,-#1+\offset) {\footnotesize{Actual impact}};
        \node[anchor=west] at (#4,-#1-\spacebetweenlines-\extralegendspacing+\offset) {\footnotesize{Future potential}};
        \draw[bar_vehicle, black!25] (#2,-#1+\offset) -- (#3,-#1+\offset);
        \draw[bar_vehicle_future, black!50] (#2,-#1-\spacebetweenlines-\extralegendspacing+\offset) -- (#4,-#1-\spacebetweenlines-\extralegendspacing+\offset);
    }
    
    \node[align=right, text width=\textboxwidth] at (\textwidth/2-\placefrommargin,0.3) {Vehicle opportunities};
    \node[align=left, text width=\textboxwidth] at (-\textwidth/2+\placefrommargin,0.3) {Bandit threats};
    \node at (0,0.3) {Sustainability domain};
    
    \draw[lines] (-\textwidth/2,0) -- (\textwidth/2,0);
    \draw[lines] (0,0) -- (0,-6);
    
    \entry{0.9}{2}{3}{0.25}{1}{Profit (Efficiency and Innovation)}
    {%
    (SDGs 8, 9)}
    {Novel products, predictive maintenance, increased efficiency and productivity}
    {Increased complexity and compliance measures could hamper product introduction}
    \entry{2.8}{1}{1.5}{1}{2}{Planet (Ecology)}
    {%
    (SDGs~6, 7~and~12-15)}
    {Early warnings for flooding risk, support biodiversity, optimized smart grid and transport operation, %
    IoT-enabled support for circular life of products}
    {Carbon footprint, increased energy demand, and usage of clean water in production of electronics and e-waste generation}
    \entry{4.7}{1}{2}{1}{2}{People (Societal sustainability)}
    {%
    (SDGs 1-3,11, 16, and~17)}
    {Precision farming, low-cost accessible e-health, broad and specialized education for all through e-learning, access to vital information, empathic smart city technology}
    {Privacy vs. surveillance, visual pollution in home, care and urban environments, digital divide increasing inequalities, cyber attacks}

    \legend{6.3}{-\textwidth/2}{-\textwidth/2+10}{-\textwidth/2+10}
    
    \end{tikzpicture}
    \caption{Could \gls{iot} technology be a vehicle or a bandit towards the SDGs? Bar lengths indicative estimates based on current information and future possibilities. %
    } %
    
    \label{fig:sdg_vehicle_or_bandit}
\end{figure*}
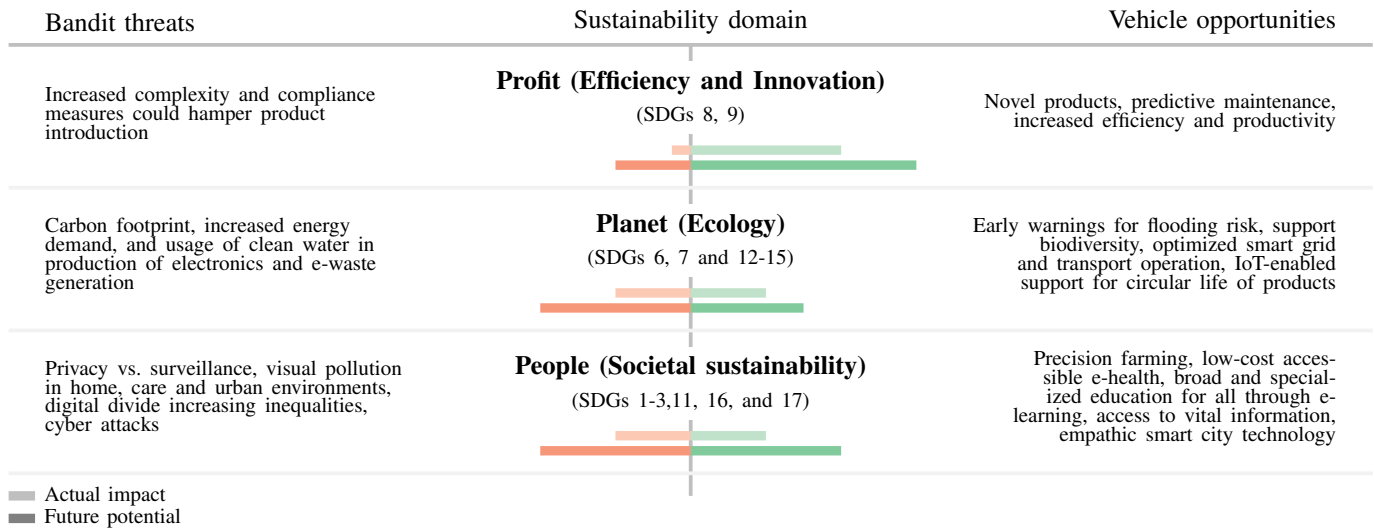

The \gls{iot} could contribute to or detract from the \glspl{sdg}. We categorize them as profit, planet, and people related, and indicate in Fig.~\ref{fig:sdg_vehicle_or_bandit} a potential for \gls{iot} to be a vehicle towards the \glspl{sdg}, and where it shows a bandit character, now or in the future. %
We first discuss the profit theme, continue with \glspl{sdg} relating to ecological impact (planet), and further come to goals relating to people's well-being.   %

\subsection{Profit: Decent work and economic growth, industry, innovation and infrastructure (SDGs~8~and~9).}

On the agenda of the 2021 World Economic Forum, \gls{iot} was a main topic and mentioned as a part of the Fourth Industrial Revolution. The technology is increasingly adopted in business environments to enable the digital revolution where a need or opportunity for decentralized data-capturing exists. \gls{iot} technology can contribute to operational excellence, tracking of equipment and goods, predictive maintenance and safety. Supply chains can be optimized, inventories can be automated, customer experiences can be improved, etc., thanks to the embedding of connected sensors. Industrial \gls{iot} saw a steep increase past the COVID-pandemic, which exposed the need to make supply chains more resilient. In many industries the development of Digital Twins is targeted to fuel innovation and efficiency. %
Clearly, \gls{iot} technology can be a strong vehicle for innovation and improving efficiency, at this moment in particular in Western countries. %

\subsection{Planet: Clean water and sanitation, climate action, life below water, life on land, affordable and clean energy, responsible production and consumption (SDGs~6, 7~and~12-15).} 

Many `smart' innovations are proposed, extending conventional systems with \gls{iot} features, with a primary goal to improve user experience, and possibly serve ecological goals. For example, in the context of the transition towards greener energy in combination with the rapidly increasing number of electric cars, NB-IoT-based systems are installed to provide real-time and location-specific data that is essential to link demand to offer. From previous sections it is clear that \gls{iot} technology comes with a direct negative impact, and as application-specific assessment ought to demonstrate a net positive environmental balance. %

\textbf{\textit{Energy consumption - Smart Lighting case}}. The smart lighting case demonstrates the importance of a quantitative assessment.  %
This analysis considers two systems: a home system and a specialized solution for street lighting in Wallonia, Belgium.
Popular smart lighting systems for home environments, such as Philips Hue, consume energy beyond the energy needed for illumination. 
Unlike standard bulbs, smart bulbs stay connected and have a sleep power of around \SI{0.4}{\watt}, resulting in an additional \SI{3.5}{kWh} per year. If a dedicated hub is required, an extra \SI{13}{kWh} per year should be accounted for~\cite{DIKEL201971}.
        The impact of the production of extra electronics would need to be included for the assessment to be complete. However, due to the long lifespan of \SI{25000}{} active hours, this impact may be negligible. The smart system could reduce the on-time per light, yet a typical \SI{10}{\watt} LED lamp would need to be off for one extra hour every day to result in a net positive environmental impact.
        In other situations, smart lighting can bring overall energy savings. \citeauthor{Pirson2023Evaluating}~\cite{Pirson2023Evaluating} present the analysis of a smart street lighting implementation. %
        The utilization phase accounts for \SI{80}{\percent} of the \gls{gwp}, and with the power of street lights typically between \SI{20} and \SI{250}{\watt}, this is an area where significant benefits can be realized by reducing operational time. %
Furthermore, it has been reported that artificial lighting disturbs wildlife and may hamper humans' sleep quality. \gls{iot}-based solutions switching off lights when safely possible, may therefore be a vehicle for different \glspl{sdg}.  %

\textbf{\textit{Broad potential of \gls{iot} for ecological benefits.}} \gls{iot}-systems can support environmental monitoring, including the state of climate change, and can contribute to a better understanding of ecological problems. 
Trees, e.g., have important functions for
carbon storage, biodiversity conservation, and people's well-being. The monitoring of heat and drought stress of trees in cities can serve as indicators of ongoing climate changes and could trigger timely action to bring water where and when needed.  %
\Gls{iot}-systems could also support forest ecosystem monitoring and management, e.g., to combat the increase of wildfires. Similarly, the observation of water quality (SDG~6) can help early detection and combating of pollution, and better preservation of wildlife (SDG~14).

The right-to-repair movement has succeeded in progressing the legal framework to encourage extended lifetimes of electronics, which is the focus of the D\textit{irective on common rules promoting the repair of goods} in the EU and the \textit{Digital Fair Repair Act} in the sate of New York. We identify an opportunity for \gls{iot} in the circular economy to manage the \gls{eol} phase. The technology could help detect and locate broken devices and e-waste.  We are, e.g., witnessing an accelerated deployment of \gls{pv} systems. These installations are bulky, not easy to recuperate and risk being left behind, while they contain (small amounts of) hazardous heavy metals. Equipped with \gls{iot} nodes for monitoring and localization of faulty panels, one could prepare for the anticipated many millions of tons of waste per year.  %

\subsection{People: No poverty, zero hunger, good health \& well-being, sustainable cities \& communities, peace and justice, strong institutions, partnerships for the goals (SDGs 1-3,11, 16, 17).} %

\textbf{No poverty, zero hunger, good health.} The goal of developing our world towards zero hunger (\gls{sdg}~2) could be served by \gls{iot} in many ways. Precision farming based on time and space specific sensor data can increase production while reducing the usage of chemicals and water.
In industrial food processing, %
the lack of appropriate control results in a lot of waste and cost ineffectiveness\rmcite{iot-food-retrofitting}.
\citeauthor{food-waste-iot}~\cite{food-waste-iot} showed that \gls{iot} can substantially reduce food waste, improve transportation and distribution efficiency, and support the quick removal of contaminated or spoiled products from the fresh food supply chain. %
An example of \gls{iot} in agriculture is its use in strawberry cultivation near M{\aa}lilla in Sweden \cite{strawberries}. Here, wireless sensors are used to remotely monitor the crop and detect conditions potentially damaging to the plants, such as frost or high temperatures, allowing mitigating measures to be taken. %
This has resulted in improvements in product quality and staff working conditions. %
However, these benefits are orthogonal to the sustainability goals, since \gls{iot} technology is used to raise the already-high quality of life and productivity in a thriving economy. The positive impact could certainly be greater if a similar system were deployed for a staple crop in an area that has low food security. Then, it could contribute to preventing food shortages as well as improving the livelihoods of the farmers, potentially lifting them out of poverty. %

The goal of good health (\gls{sdg}~3) can benefit from \gls{iot} solutions through real-time patient monitoring and remote support, for example for smart inhalers. %
In \glspl{wban}, sensors are connected to capture measurements on different parts of the body. %
Physiotherapists can benefit from multi-sensor measurements, e.g., using \glspl{imu} to study the long-term evolution of symptoms, instead of sporadically using expensive equipment.

\textbf{Sustainable cities (SDG~11).} %
Cities contribute to \SI{70}{\percent} of global carbon emissions, \SI{80}{\percent} of energy consumption and has a settlement for more than \SI{60}{\percent} of the global population by 2030. %
Even if cities are at the center of a discussion between scientists and urban planners as a major cause of environmental crises, the focus is mostly on solving the consequences of these crises instead of preventing them.  
Cities' transformation should consider new relationships with nature and provide transparent information to citizens about data capturing. \Gls{iot} technology can serve cities, from improving efficiency to monitoring critical situations such as floods or heatwaves.
However, many \gls{iot} systems have questionable value for achieving these challenges, as a lot of data is captured by big tech companies with commercial motives.
Technology deployment needs to be reoriented to contribute to the \glspl{sdg}. 
Moreover, city planners and designers should investigate the impact of the appearance of ICT technology:
should we hide visually disturbing electronics, or create aesthetically attractive designs in dialogue with citizens to stimulate empathy for technology?

\textbf{Equal opportunities: decrease inequalities, gender gap, and education for all (SDGs~4, 5 and 10).} 
\Gls{iot}-based digitization of services could decrease inequalities, for example e-health applications could lower cost of medical care.  In practice, however, it so far is mostly a vehicle for the rich and educated, and digital literacy has become a source of inequality. As an example, regions in Africa where resources are extremely scarce could benefit greatly from `smart farming'.  Currently, they to a large extent lack both the expertise and the investment capabilities to adopt this technology. %
The digital divide also has a gender dimension. Data\rmcite{DIgnazio2020-DF-10}, and in particular the interpretation thereof, is not neutral, and biases may be reinforced. It has been advocated that `the age of data' opens an opportunity to change and break traditional inequalities.

\textbf{Human rights: Peace and justice, strong institutions, partnerships for the goals (SDGs 1, 16, and~17).} Knowledge itself is power. Monitoring and other data generated by \gls{iot} technologies can be used and misused by authorities and individuals. For example, the fact that large sets of internet-connected PV installations could be simultaneously switched off remotely by other entities than the owners, creates a threat to a critical resource.
Digital technologies in cities and home appliances are often invisible to citizens and users or they create visual pollution. They can increase comfort and safety, yet also raise a distrust and fear of being monitored. There is a lack of transparency: who owns the data? \Gls{iot}-technology deployment should get support from strong institutions to ensure human rights are respected, while vice versa they could also benefit from access to clear and unbiased data. %

\section{Conclusions and suggestions}\label{sec:concl}

(Io)Things are not always what they seem. The introduction of IoT technology has been presented as a vehicle towards the \glspl{sdg}. However, we showed that its bandit trait is more sure when the ecological and people dimension are considered. %
The introduction of NB-IoT in radio networks induces a significant energy penalty. \Gls{iot} devices come with a carbon footprint that will grow with a massive deployment, and the production of the electronics consumes (rare) materials and clean water. The need for specific quantitative studies to assess the net energetic impact was illustrated.  %
We observe that for many applications, the jury is still out on whether the overall net effect of \gls{iot} deployment will be positive. %

Based on these insights, we suggest R\&D directions for \gls{iot} on the road towards the \glspl{sdg} on wireless technologies, hardware design, and applications. %

\textbf{Efficient IoT support in future networks.} 
Promising technologies include crowdsourced networks, where operators can coordinate organically constructed user-centric wireless networks. \Gls{cf} networking principles bear an interesting potential for selective activation of network segments based on \gls{iot} requirements and their location. The support for critical \gls{iot} services hinders energy savings in the network. There is a need to explore strategies that decouple network functions and signals.

\textbf{Towards circular \gls{iot} nodes.} %
New design paradigms should prepare \gls{iot} devices for repair, reuse, and recycle. These can include adopting a modular approach, strategies to extend the nodes' lifetime, and  %
the usage of biodegradable technologies.

\textbf{Backcasting approach for \gls{iot} systems.} We endorse the recommendations, to perform a quantitative assessment of the environmental impact of \gls{iot} technology in each use case, `to define a vision of a desirable future and then work backward from the end-point vision to the present'~\cite{Pirson2023Evaluating}\rmcite{Rasoldier2022How}.
\Gls{iot} can be a game changer to address sustainability problems. The unexplored frontiers of multidisciplinary collaborations and creative designs can prosper to power the vehicle of sustainable \gls{iot} technology, and minimize its inevitable banditry. %

\section*{Acknowledgements}
The authors are thankful to many contributors, in particular Pål Frenger, Jarne Van Mulders, Sofya Ignatenko, David Bol, Stien Dethier, and Dimitri Coppens. This research was supported by the REINDEER project funded by the European Union’s Horizon 2020 research and innovation programme under grant agreement No. 101013425 and the FWO sabbatical program grant K801323N.

\vskip -1.75\baselineskip plus -1fil
\begin{IEEEbiographynophoto}{Jona~Cappelle}
(IEEE Member, jona.cappelle@kuleuven.be) earned his M.Sc. in Engineering Technology from KU Leuven, Belgium in 2020. Currently, he is a PhD researcher at the DRAMCO research group. He is working on architectural concepts and design for circular IoT. His main interests are low-power embedded systems, wireless communications, and sustainable IoT.

\end{IEEEbiographynophoto}%
\vskip -2.75\baselineskip plus -1fil
\begin{IEEEbiographynophoto}{Liesbet~Van~der~Perre} (IEEE senior member, liesbet.vanderperre@kuleuven.be) received the M.Sc. and Ph.D. in electrical engineering from KU Leuven, Belgium, in 1992 and 1997. She was with the research institute imec from 1997 till 2015. Since 2016, she
is a professor at KU Leuven and a guest professor at ULund, Sweden, focusing on sustainable IoT and wireless systems.
\end{IEEEbiographynophoto}%
\vskip -2.75\baselineskip plus -1fil
\begin{IEEEbiographynophoto}{Emma~Fitzgerald}
(IEEE Member, emma.fitzgerald@eit.lth.se) received the B.Sc., B.E., and Ph.D. degrees from The University of Sydney, Australia, in 2008 and 2013. 
She was Associate Professor with the EIT Department at ULund, Sweden, and a Researcher with the Warsaw University of Technology, Poland. She focuses on network performance analysis, IoT, and beyond 5G.
Currently, she works as a software engineer at Atlassian.
\end{IEEEbiographynophoto}%
\vskip -2.75\baselineskip plus -1fil
\begin{IEEEbiographynophoto}{Simon~Ravyts}
(IEEE Member, simon.ravyts@kuleuven.be) received the M.Sc. degree in energy and electrical and electrical power engineering from KU Leuven and UGent, Belgium, in 2014 and 2016. He received his Ph.D. from KU Leuven. He is currently a postdoc at ELECTA, KU Leuven. 
His research interests include integrated photovoltaics modules, power electronics and low-voltage DC power distribution networks.
\end{IEEEbiographynophoto}%
\vskip -2.75\baselineskip plus -1fil
\begin{IEEEbiographynophoto}{Weronika~Gajda}
(weronika.gajda@kuleuven.be) is an interdisciplinary PhD researcher at DRAMCO, ESAT and faculty of architecture, KU Leuven, Belgium.
Her research focuses on how media ecologies are reshaping both the architectural practice and collective experiences within public spaces.
\end{IEEEbiographynophoto}%
\vskip -2.75\baselineskip plus -1fil
\begin{IEEEbiographynophoto}{Valentijn De Smedt}
(IEEE Member, valentijn.desmedt@kuleuven.be) received the M.Sc. and the Ph.D. degree in electrical engineering from KU Leuven, Belgium, in 2007 and 2014 respectively.
He has since worked at MICAS and MinDCet and is now a professor at the ADVISE research group, at ESAT, KU Leuven, Belgium, working on radiation-tolerant control systems for power and sensing applications.
\end{IEEEbiographynophoto}%
\vskip -2.75\baselineskip plus -1fil
\begin{IEEEbiographynophoto}{Bert~Cox}
(IEEE Member, bert.cox@kuleuven.be) earned his M.Sc. and Ph.D. in Engineering Technology from KU Leuven, Belgium, in 2015 and 2022. Currently, he is a postdoc in the research group DRAMCO. His ongoing efforts are in the field of acoustic signalling, low-power design and embedded communication.
\end{IEEEbiographynophoto}%
\vskip -2.75\baselineskip plus -1fil
\begin{IEEEbiographynophoto}{Gilles~Callebaut} (IEEE Member, gilles.callebaut@kuleuven.be) earned his M.Sc. and Ph.D. in Engineering Technology from KU Leuven, Belgium, in 2016 and 2021. 
\arxiv{His doctoral research focused on single and multi-antenna technologies to reduce energy consumption in IoT devices.}
Currently, he is a postdoc in the research group DRAMCO. He advocates for sustainable, low-power communication systems and promotes scalable, low-carbon technologies in IoT and 6G.
\end{IEEEbiographynophoto}%

\printbibliography

\end{document}